\documentclass[conference]{IEEEtran}
\IEEEoverridecommandlockouts
\usepackage{cite}
\usepackage{amsmath,amssymb,amsfonts}
\usepackage{algorithmic}
\usepackage{graphicx}
\usepackage{textcomp}
\usepackage{xcolor}
\usepackage{url}
\usepackage{booktabs}
\usepackage{multirow}
\usepackage{pifont}
\usepackage{balance}
\def\BibTeX{{\rm B\kern-.05em{\sc i\kern-.025em b}\kern-.08em
    T\kern-.1667em\lower.7ex\hbox{E}\kern-.125emX}}

\def\W{{\mathbf W}}
\def\x{{\mathbf x}}
\def\ii{{\hat{\imath}}}	
\def\ij{{\hat{\jmath}}} 
\def\ik{{\hat{\kappa}}}	

\begin{document}

\title{PHemoNet: A Multimodal Network for Physiological Signals\\
\thanks{This work was partially supported by the Italian Ministry of University and Research (MUR) within the PRIN 2022 Program for the project ``EXEGETE: Explainable Generative Deep Learning Methods for Medical Signal and Image Processing", under grant number 2022ENK9LS, CUP B53D23013030006, and in part by the European Union under the National Plan for Complementary Investments to the Italian National Recovery and Resilience Plan (NRRP) of NextGenerationEU,  Project PNC 0000001 D3 4 Health - SPOKE 1 - CUP B53C22006120001.}
}

\author{\IEEEauthorblockN{Eleonora Lopez, Aurelio Uncini and Danilo Comminiello}
        \IEEEauthorblockN{\textit{Dept. Information Engineering, Electronics and Telecommunications (DIET), Sapienza University of Rome, Italy}\\
        Email: eleonora.lopez@uniroma1.it.}
}

\maketitle

\begin{abstract}

Emotion recognition is essential across numerous fields, including medical applications and brain-computer interface (BCI). Emotional responses include behavioral reactions, such as tone of voice and body movement, and changes in physiological signals, such as the electroencephalogram (EEG). The latter are involuntary, thus they provide a reliable input for identifying emotions, in contrast to the former which individuals can consciously control. These signals reveal true emotional states without intentional alteration, thus increasing the accuracy of emotion recognition models. However, multimodal deep learning methods from physiological signals have not been significantly investigated. In this paper, we introduce PHemoNet, a fully hypercomplex network for multimodal emotion recognition from physiological signals. In detail, the architecture comprises modality-specific encoders and a fusion module. Both encoders and fusion modules are defined in the hypercomplex domain through parameterized hypercomplex multiplications (PHMs) that can capture latent relations between the different dimensions of each modality and between the modalities themselves. The proposed method outperforms current state-of-the-art models on the MAHNOB-HCI dataset in classifying valence and arousal using electroencephalograms (EEGs) and peripheral physiological signals. The code for this work is available at \url{https://github.com/ispamm/MHyEEG}.

\end{abstract}


\section{Introduction}
\label{sec:intro}

Hypercomplex neural networks represent a sophisticated class of neural models that extend traditional real-valued networks into higher-dimensional spaces using hypercomplex algebras, such as complex numbers, quaternions, and octonions. These networks exploit the inherent mathematical properties of hypercomplex numbers to model complex relationships within data, providing a more powerful mechanism for capturing both local and global patterns \cite{comminiello2024demystifying}. Parameterized Hypercomplex Neural Networks (PHNNs) further extend this capability by a parameterization through a hyperparameter $n$ which enables them to simulate different hypercomplex domains such as complex and quaternion spaces \cite{zhang2021phm, grassucci2022phnns}. This adaptability results in significant parameter efficiency and improved data representation, making PHNNs particularly suitable for tasks involving multidimensional inputs. Overall, hypercomplex layers offer a promising approach to advancing deep learning models by leveraging the unique advantages of hypercomplex algebra.

For these reasons, hypercomplex architectures have been developed for medical applications \cite{basso2023ecg}, as medical data often present multiple modalities or views that can be effectively exploited through an ad hoc hypercomplex architecture thanks to the properties of hypercomplex algebra \cite{lopez2022multi}. Specifically, a recent work has introduced a multimodal network for emotion recognition from physiological signals, featuring a hypercomplex fusion module \cite{lopez2023hypercomplex}. This module enhances fused representations by leveraging the relations among different modalities using hypercomplex algebra. It begins to address challenges in the literature, such as extensive preprocessing requiring domain-specific knowledge \cite{nakisa2020automatic}, which prevents neural networks from learning important features from raw data, and the use of single modalities when emotional responses are intrinsically multimodal \cite{zhang2022multimodal}. However, this approach still faces issues, mainly due to overfitting. Additionally, the current architecture employs hypercomplex layers only at the fusion step, while the encoders remain in the real domain. Incorporating hypercomplex layers in the encoders could further exploit the benefits of hypercomplex algebras, potentially improving the network performance and robustness.

Therefore, in this paper, we propose a fully hypercomplex network (PHemoNet) comprising both encoders and fusion module in the hypercomplex domain. In detail, the encoders and the fusion module are composed of parameterized hypercomplex multiplications (PHMs). In this manner, at encoder level the model learns enhanced modality-specific embeddings thanks to hypercomplex algebra. In fact, each encoder specific to the different input physiological signal is defined in a different hypercomplex domain that is the natural domain of the signal, i.e., its original dimensionality. We employ as input signals the electroencephalogram (EEG), eye data, galvanic skin response (GSR), and electrocardiogram (ECG), as all of them reflect emotional changes \cite{soleymani2011mahnob}. Then, a refined and more efficient hypercomplex fusion module learns the fused representation from the single modality embeddings. We evaluate our fully hypercomplex approach on the MAHNOB-HCI \cite{soleymani2011mahnob} benchmark for arousal and valence classification and compare it against previous state-of-the-art networks. We find that the proposed method surpasses the results in the literature on both tasks, advancing the state-of-the-art.

The rest of the paper is organized as follows. In section \ref{sec:related} we discuss the recent works on hypercomplex models and emotion recognition. In Section~\ref{sec:background} we provide background on the theory behind hypercomplex networks, while we describe the details of the proposed architecture in Section~\ref{sec:method}. In Section \ref{sec:experiments} we present the experimental results from the conducted evaluation. Finally, we outline our conclusions in Section \ref{sec:conclusion}.

\section{Related works}
\label{sec:related}

\subsection{Hypercomplex learning}
\label{sec:related_hyp}

Hypercomplex models are spreading across the research community unifying fields such as mathematics, signal processing, and deep learning. Starting from studying mathematical properties of different subdomains, including the widely studied quaternion domain \cite{zheng2022quaternion, qin2022fast, vieira2023dual}, along with the tessarine \cite{navarro2020tessarine} and complex \cite{bassey2021survey} domains, to exploiting their properties and capabilities tailoring models to specific applications. Indeed, QNNs excel in learning representations of multi-dimensional data such as audio \cite{brignone2022efficient, grassucci2023dual}, speech \cite{muppidi2021speech, guizzo2023learning}, and images \cite{frants2023qcnn, parcollet2019image}. Even with medical images which are typically grayscale and are therefore 1-dimensional, quaternion algebra has been exploited by integrating the model with wavelet transforms \cite{sigillo2023generalizing, grassucci2023grouse}.
However, algebras follow the Cayley-Dickson system, i.e., there exists an algebra only for $2^m$ where $m=1,2,\cdots$. Thus, PHNNs have been developed and have overcome the limitations imposed by specific hypercomplex domains by learning the algebra directly from the data \cite{zhang2021phm, grassucci2022phnns, le2021parameterized}. In this way, the capabilities of hypercomplex models can be brought to domains for which an algebra does not yet exist. For example, images can be processed in their natural 3D domain \cite{grassucci2022i2i}, and similarly, audio \cite{panagos2022compressing} and text \cite{sfikas2023shared} data can also be handled in their natural domains. Moreover, their flexibility allows to design architectures for multi-view \cite{lopez2022multi} or multimodal \cite{lopez2024attention} data that is typical in medical scenarios.  This versatility also enables the development of lightweight models, a crucial property for medical applications like detecting atrial fibrillation from ECG signals collected by wearable devices \cite{basso2023ecg}. Finally, owing to their advantages, numerous studies are now delving into their theoretical aspects, ranging from initialization techniques \cite{mancanelli2023phydi} to studying their learning behavior by rendering them directly interpretable \cite{lopez2024towards}.


\subsection{Emotion recognition}

Given the reliability of physiological signals for the task of emotion recognition, several works developed machine learning-based methods, such as emotion recognition from photoplethysmography (PPG) and GSR \cite{dominguez2020machine} and valence recognition from EEG extracted features \cite{abdel2023efficient}. Moreover, recent studies developed deep learning-based techniques, such as an attention-based hybrid model \cite{zhang2023attention} and a heterogenous convolutional neural network (CNN) with multimodal factorized bilinear pooling \cite{zhang2022emotion} for EEG emotion recognition. The main issue with all aforementioned works lies in the extensive preprocessing and domain knowledge required. Indeed, they require hand-crafted features typical of pure machine learning methods \cite{dominguez2020machine, abdel2023efficient}, or power spectral density (PSD) and differential entropy (DE) \cite{zhang2023attention}, or a study of the impact of the frequency bands on the downstream task \cite{zhang2022emotion}. In addition to preprocessing, a vast majority focuses on a single-modality approach, while humans express emotions in a multimodal manner. Recent multimodal approaches for emotion recognition include a temporal multimodal model from EEG and blood volume pulse (BVP) \cite{nakisa2020automatic} and a cross-subject method based on completeness-induced adaptative broad learning from EEG and eye movement signals \cite{gong2024ciabl}. Finally, in our previous work \cite{lopez2023hypercomplex} we introduced HyperFuseNet, a multimodal architecture that learns directly from raw signals. In this paper, we extend this and develop a fully hypercomplex model to overcome its generalization issue.

\section{Background}
\label{sec:background}

\begin{figure*}[t]
    \centering
    \includegraphics[width=\textwidth]{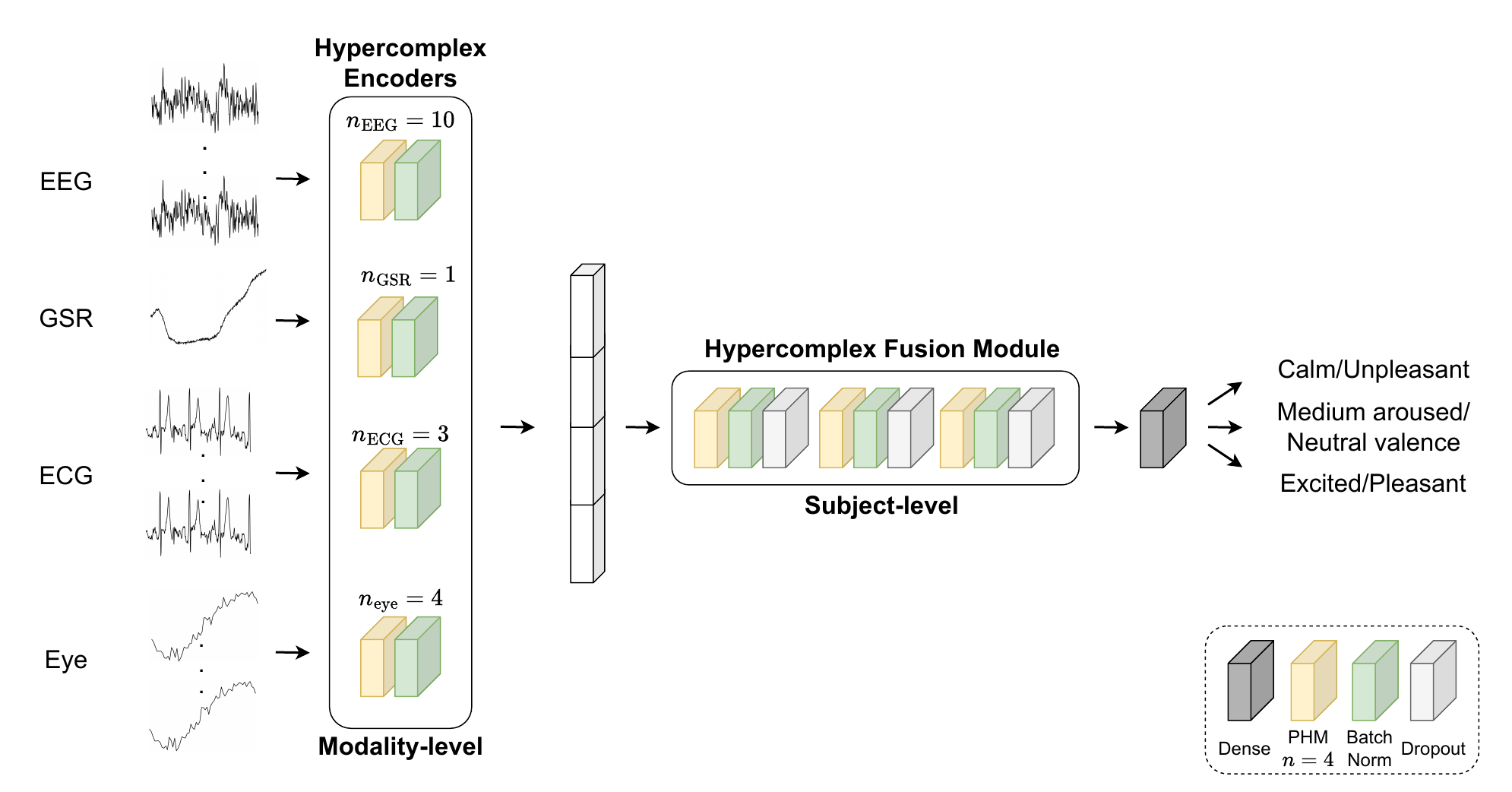}
    \caption{PHemoNet. Fully hypercomplex architecture with encoders operating in different hypercomplex domains according to $n$ and a refined hypercomplex fusion module with $n_{\text{fusion}}=4$. Finally, a classification layer yileds the final output for valence/arousal prefiction.}
    \label{fig:method}
\end{figure*}

Hypercomplex models extend the conventional real-valued neural network paradigms by leveraging hypercomplex number systems, such as quaternions and octonions. These systems, which generalize complex numbers, enable more compact and efficient representations of data, especially in applications involving high-dimensional spaces and correlated features \cite{comminiello2024demystifying}. Hypercomplex numbers extend complex numbers by incorporating additional dimensions. For instance, quaternions, a four-dimensional number system, consist of one real part and three imaginary parts. Mathematically, a quaternion $q$ can be represented as $q = q_0 + q_1 \ii + q_2 \ij + q_3 \ik$, with $q_i \in \mathbb R$ and $\ii$, $\ij$ and $\ik$ are imaginary units satisfying specific multiplication rules. Hypercomplex neural networks utilize these properties to construct layers that perform operations in the hypercomplex domain. A common approach is to replace real-valued weight matrices with hypercomplex-valued ones, resulting in models that can capture intricate correlations in the data with fewer parameters \cite{comminiello2024demystifying}.

Parameterized hypercomplex networks (PHNNs) \cite{zhang2021phm, grassucci2022phnns} introduce additional flexibility by allowing the network to learn optimal parameters that define the hypercomplex algebra used in the model. This approach generalizes fixed hypercomplex models like quaternion and octonion networks by incorporating trainable parameters that adjust the multiplication rules and interactions between different components of the hypercomplex numbers. In a parameterized hypercomplex neural network, the weight matrix $\W$ is not strictly defined by the fixed rules of quaternion or octonion algebra. Instead, it is parameterized by a set of learnable parameters leveraging Kronecker products:

\begin{equation}
    \W = \sum_{i=1}^n \mathbf{A}_i \otimes \mathbf{F}_i.
\label{eq:matirx}
\end{equation}

\noindent Herein, the matrices $\mathbf{F}_i$ serve as the conventional learnable weights for a specific layer, while matrices $\mathbf{A}_i$ encode the algebra rules. The hyperparameter $n \in \mathbb{N}$ defines the dimensionality of the number system used by the model, for instance, $n=4$ corresponds to a Quaternion Neural Network (QNN). Following this definition of the weight matrix, a parameterized hypercomplex multiplication (PHM) layer is defined as its real-valued counterpart:

\begin{equation}
    \mathbf{y} = \text{PHM}(\x) = \W \cdot \x + \mathbf{b},
\label{eq:phm_phc}
\end{equation}

\noindent where $\W$ is defines according to eq.~\ref{eq:matirx}. Thanks to this formulation, PHM layers and consequently PHNNs can operate also on domains for which an algebra does not yet exist, as it is encoded by $\mathbf{A}_i$ during training. Moreover, they preserve the advantages of QNNs, i.e., efficiency and the ability to capture correlations. 


\section{Proposed Method}
\label{sec:method}

We propose a fully hypercomplex network (PHemoNet) for multimodal emotion recognition, which we show in Fig. \ref{fig:method}. The architecture comprises both encoders and fusion module defined in different hypercomplex domains. The encoders learn an embedding specific to the modality, which will then be fused with the other modalities embeddings by the fusion module. In fact, the latter learns a fused representation which allows to leverage cross-modality information for a better classification output. Each input signal is processed by a specific encoder defined in the original domain of the physiological signal by setting the hyperparameter $n$ of the PHM layers. In detail, for GSR we set $n_{\text{GSR}}=1$ since it is a 1-dimensional signal, for ECG we configure $n_\text{ECG}=3$, for the eye signals we set $n_{\text{eye}}=4$ since it comprises four dimensions including gaze coordinates, pupil dimensions, and eye distances, and finally, we set $n_{\text{EEG}}=10$ since we select ten electrodes. Each encoder is composed of one PHM layer with 128, 131, 1020, and 513 hidden units for eye signals, GSR, EEG, and ECG respectively, and interleaving batch normalization and ReLU activation functions. Then the refined fusion module is composed of three PHM layers with $n_{fusion}=4$ in accordance with the number of input modalities, and with an initial 1792 hidden units which are halved by each PHM layer. The same interleaving operations as in the encoders are applied with the addition of dropout layers with a rate of 0.5. In contrast with the fusion module in \cite{lopez2023hypercomplex}, which had only one dropout layer with a lower probability and had an additional PHM layer which contributed to the overfitting behavior of the model. A final classification layer yields the predicted class, i.e., calm, medium aroused, excited for arousal and unpleasant, neutral valence, and pleasant for valence. In addition to the refined fusion module, the main advantage of PHemoNet is the encoders operating in hypercomplex domains. In this way, the final architecture turns out to be more lightweight given the reduction of parameters brought by PHM layers \cite{zhang2021phm} and the ability to capture relations among channels of each input signal thanks to hypercomplex algebra properties.

\section{Experimental results}
\label{sec:experiments}

\begin{table*}[t]
\centering
\caption{Results on MAHNOB-HCI of PHemoNet compared against two state-of-the-art architectures. Results in bold and underlined correspond to the best and second best, respectively.}
\label{tab:results}
\begin{tabular}{@{}lccclcc@{}}
\toprule
\multirow{2}{*}{Model}  & \multicolumn{2}{c}{Arousal} &  & \multicolumn{2}{c}{Valence} \\ \cmidrule(lr){2-3} \cmidrule(l){5-6} 
                                            & F1-score     & Accuracy     &  & F1-score     & Accuracy     \\ \midrule
Dolmans \cite{dolmans2021workload} & 0.389 $\pm$ 0.011 & 40.90 $\pm$ 0.62  &  & 0.383 $\pm$ 0.012  & 40.24 $\pm$ 1.04  \\
HyperFuseNet \cite{lopez2023hypercomplex} & \underline{0.397} $\pm$ 0.018  & \underline{41.56} $\pm$ 1.33  &  & \underline{0.436} $\pm$ 0.022  & \underline{44.30} $\pm$ 2.01  \\ 
PHemoNet (ours) & \textbf{0.401} $\pm$ 0.022 & \textbf{42.54} $\pm$ 1.98 & & \textbf{0.505} $\pm$ 0.005 & \textbf{50.77} $\pm$ 0.50\\
\bottomrule
\end{tabular}
\end{table*}

\subsection{Dataset and preprocessing}

The dataset we employ for our experiments is the MAHNOB-HCI database for affect recognition \cite{soleymani2011mahnob}. It provides different physiological signals, video, and audio recordings of 27 subjects during an emotional experiment consisting of the view of video clips. Each recording is annotated with labels for arousal, i.e., calm, medium aroused, and excited, and valence, i.e., unpleasant, neutral, and pleasant. As already mentioned, for our experiments we utilize the eye data, which include eye distances, pupil dimensions and gaze coordinates, GSR, ECG, and EEG. 

Regarding preprocessing and data augmentation, we apply the same operations as in \cite{lopez2023hypercomplex}. For eye signals, we take the average of the left and right eye and do not remove -1 values as they are related to blinks which could be part of the emotional response. For EEG signals we select the ten electrodes most connected to emotions, i.e.,  F3, F4, F5, F6, F7, F8, T7, T8, P7, and P8 \cite{topic2022reduced, msonda2021channel}, and we reference it with respect to the average. EEG, GSR, and ECG are resampled from 256Hz to 128Hz. We apply a band-pass filter of 0.5-45Hz for ECG signals, of 1-45Hz for EEG, and a low-pass filter at 50Hz for GSR. All of them are then filtered with a notch filter at 50Hz. Additionally, we correct the baseline of GSR to account for its initial offset as in \cite{lopez2023hypercomplex}. The signals are augmented with the addition of Gaussian noise and scaling operations as in \cite{lopez2023hypercomplex}. Each sample is a 10s segment of the original 30s recordings, which are split into training and testing in a stratified fashion, taking 80\% and 20\%, respectively.

\subsection{Experimental setup}

We employ accuracy and F1-score as metrics. The F1-score is defined as the harmonic mean of recall and precision metrics, thus it accounts for the imbalance present in the input data.
The models are trained with a cross-entropy loss and Adam optimizer for 50 epochs with early stopping and the patience hyperparameter set to 10 in accordance with the F1-score. The loss is updated following a one-cycle policy configured with $7.96 \times 10^{-6}$ as maximum learning rate, 10 as dividing factors, 0.425\% as the percentage of increasing steps, 0.7985 as maximum momentum, 0.7403 as minimum momentum and a linear annealing strategy.

\subsection{Results and discussion}

We report the results of the experimental evaluation in Tab.~\ref{tab:results}. We compare the proposed method against two state-of-art approaches \cite{lopez2023hypercomplex, dolmans2021workload}. It is clear that the proposed fully hypercomplex architecture allows to generalize better and thus outperforms previous methods for both arousal and valence classification. Indeed, the generalization ability is due to several advantages of PHemoNet. Firstly, being a fully hypercomplex architecture, there is a greater reduction in parameters compared to HyperFuseNet, which only had the fusion module in the hypercomplex domain. This allows to build a model that is better tailored to the task instead of being overparameterized. Secondly, thanks to a refined fusion module with less layers and more dropout operations, it allows to not overfit during the fusion process. Finally, but most importantly, the encoders are able to process each input signal in its natural number domain by defining them in specific hypercomplex domains, and thus learn a better representation of the different modalities. These advantages yield an F1-score of 0.401 for arousal classification and 0.505 for valence classification.

\section{Conclusion}
\label{sec:conclusion}

We have proposed a fully hypercomplex network, PHemoNet, for multimodal emotion recognition. Different physiological signals are processed by modality-specific encoders in their natural domain by setting the hyperparameter $n$ of PHM layers as explained in Section~\ref{sec:method}. Moreover, we revise the fusion module previously proposed in \cite{lopez2023hypercomplex} in order to not make the final architecture overfit. With the main advantage of a fully hypercomplex multimodal network, PHemoNet is able to generalize better with respect to state-of-the-art models and thus outperform them. In future works, we aim to investigate convolution operations at encoder level with parameterized hypercomplex convolutional (PHC) layers \cite{grassucci2022phnns}.

\balance
\bibliographystyle{IEEEtran}
\bibliography{biblio}

\end{document}